# PHASE ROTATION OF MUON BEAMS FOR PRODUCING INTENSE LOW-ENERGY MUON BEAMS *

D. Neuffer, Fermilab, Batavia IL 60532, USA
Y. Bao and G. Hansen, UC Riverside, CA 60439, USA

*Abstract*

Low-energy muon beams are useful for rare decay searches, which provide access to new physics that cannot be addressed at high-energy colliders. However, muons are produced within a broad energy spread unmatched to the low-energy required. In this paper we outline a phase rotation method to significantly increase the intensity of low-energy muons. The muons are produced from a short pulsed proton driver, and develop a time-momentum correlation in a drift space following production. A series of rf cavities is used to bunch the muons and phase-energy rotate the bunches to a momentum of around 100 MeV/c. Then another group of rf cavities is used to decelerate the muon bunches to low-energy. This obtains ~0.1 muon per 8 GeV proton, which is significantly higher than currently planned Mu2e experiments, and would enable a next generation of rare decay searches, and other intense muon beam applications.

## INTRODUCTION

Mu2e and other rare decay experiments require low-energy μ beams. A proton beam on a production target produces a broad spectrum of secondary particle energies, and the experiments have selected particles from the low energy tail of production, $P_\mu < 100$ MeV/c (see Fig. 1). Many more μ's are produced in the 100-500 MeV/c range. High-intensity high-energy μ applications have developed methods for collecting and cooling μ's from that production maximum. In this paper we initiate adapting these methods toward collection of μ's from this broader production peak and decelerating them into the low-energy beams needed for rare decay searches.

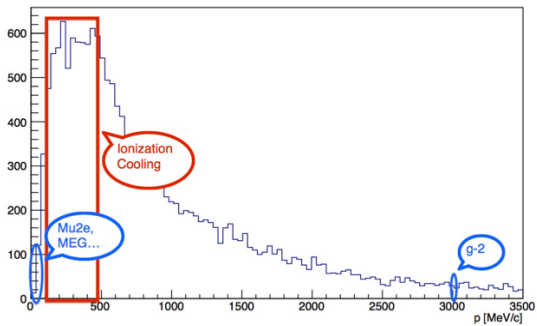

Figure 1: Muon momentum spectrum produced by 8 GeV protons.

## COLLECTION SCENARIO

For high-energy applications, beam collection schemes oriented toward the collection of 100—400 MeV/c π's to obtain μ's were developed, as shown in figure 2. A high intensity proton beam bunch is incident on a target, producing a broad spectrum of secondary π's. These are captured within a high-field solenoid and directed into a drift transport. π decay within the drift produces μ's, with a time-momentum correlation. A buncher rf system forms the μ's into bunches, and a subsequent system phase-energy rotates the bunches to ~equal energies. (Scenarios matched to rf frequencies of ~325MHZ and ~200 MHz were developed.) The μ bunches are then cooled and accelerated for HEP applications. The system was designed to accumulate ~250 MeV/c μ's, which is optimal for production and cooling (see fig. 1). The front end can be adapted for low-energy applications; the simplest method would be to use the initial system, with cooling, and decelerate the μ's into the required momentum regime. This would work, but requires a front end system that is much more expensive than the more limited potential resources of low-energy experiments. In the present paper we explore adapting these techniques within much more restricted parameters. Large intensity increases over the present low-energy capture are possible.

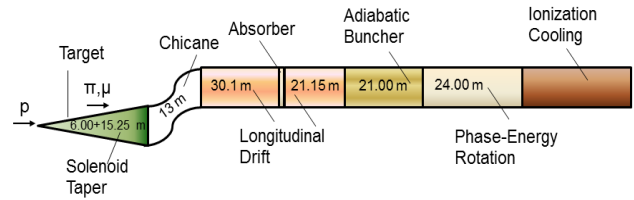

Figure 2: Overview of a front end for HEP μ applications.

## LOW-ENERGY SCENARIOS

The scenario described above is initiated by single intense proton bunches on the target. The JPARC accelerators can provide beam precisely matched to that requirement. Other facilities (cw linacs, or high-frequency synchrotrons) would require proton bunching systems. A possible layout for a low energy muon source facility is shown in figure 3.

Adaptation toward a lower energy source implies collection at lower energy. We consider collection of π→ μ at ~70—200 MeV/c, with the phase-energy rotation obtaining bunches at 100 to 150 MeV/c. At these energies the length of the system can be reduced by ~1/2 or more, with reduced rf requirements.

Also the cooling system used for HEP applications is not included. Cooling is inefficient at lower momentum and would require collection and cooling at 200—250 MeV/c with subsequent deceleration. The low energy collection will still lead to greatly increased μ intensities.

___________________________________________

* Work supported by by FRA Associates, LLC under DOE Contract No. DE-AC02-07CH11359.
#neuffer@fnal.gov

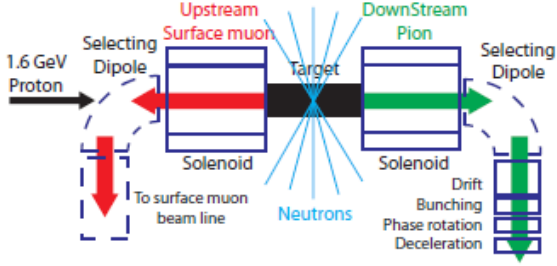

Figure 3: Overview of a general purpose muon source. [from ref. 3]. Backward production provides surface muons; forward produced π's can be transported into the phase rotation system to produce a high-intensity low-energy μ beam.

*Sample Scenario*

As a first scenario, we use a simple extrapolation of the HEP muon source, consisting of a drift, buncher, and rotator, with each of those scaled to less than half the size of the HEP source. Fig. 4 displays a conceptual layout of the source. It consists of a proton beam on a target producing π's with a collection lens matched into a drift, where the π's decay into μ's. The drift is ~ 20.4 m long and is followed by buncher and rotator sections that include high frequency rf. In the buncher section the rf frequencies decrease from 507 MHz to 363 MHz while the gradients ramp up from 0 to 15 MV/m. (The rf has 2 0.25m cavities in 0.75m long cells.) In the rotator, the gradient is fixed at 15MV/m and the frequency decreases from 360 MHz to 314 MHz. 2T solenoids are used for transverse focusing throughout the drift, buncher and rotator.

The buncher is designed to capture 100—150 MeV/c μ's in a train of bunches of different energies. The rotator moves the bunches to equal energies ($P_\mu$= 112 MeV/c or $E_\mu$= 48.3MeV).

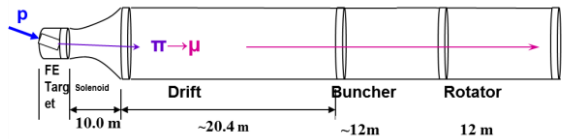

Figure 4: Conceptual Layout of Front End for low energy μ capture.

*Simulation Results*

The front end has been simulated using the ICOOL simulation program [5], with initial beam generated at the target using MARS[4] for the MAP front end, assuming an 8 GeV input proton beam.

In an initial simulation, ~0.08 μ/p are captured within ~30 bunches (~30m long bunch train) with a mean momentum of ~120 MeV/c. The p-cτ distribution of the beamV at the end of the front end is displayed in figure 5.

This initial case was motivated by following the MAP HEP front end design, which was matched to the 325 MHz front end. A somewhat better match is obtained by matching to ~200 MHz and reducing gradients to 10—12 MV/m. Then the rf frequencies range from 338 to 210 MHz. The rf cavities have a larger acceptance, corresponding to the initial beam size, and the gradients match those established in operation in the MICE MTA experiment.

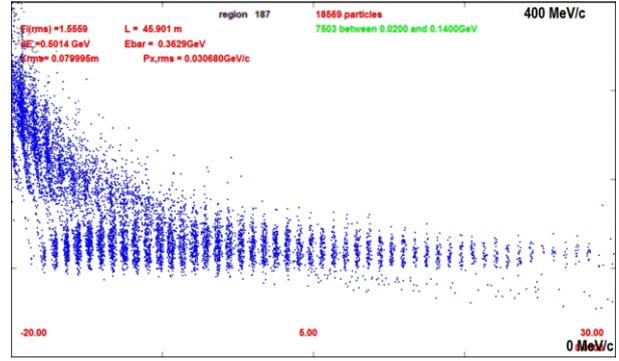

Figure 5: cτ- P distributions of the beam at the end of the Rotator (314MHz example). Vertical scale is 0 to 400 MeV/c; horizontal scale covers 50m.

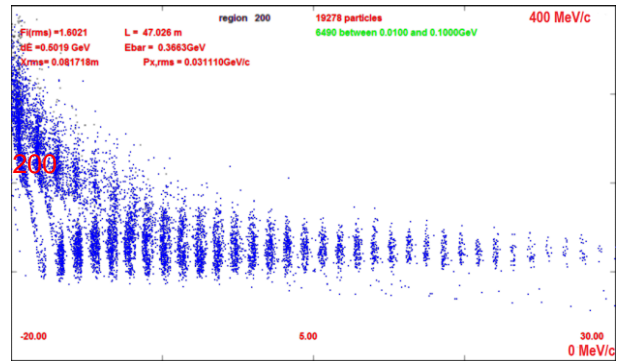

Figure 6: cτ- P distributions of the beam at the end of the Rotator (210MHz example). Vertical scale is 0 to 400 MeV/c; horizontal scale covers 50m.

*Deceleration scenarios*

The ~100 MeV/c beam is not low enough for some applications. The buncher must then be followed by deceleration. The simplest decelerator is a constant-frequency rf system matched to the bunch spacing at the end of the rotator (~208.7MHz), with phase matched to reduce the energy at ~5MV/m. A first attempt is relatively efficient in reducing the the beam momentum to < 70MeV/c (21MeV), as may be needed for a rare decay experiment.

Roughly half of the trapped μ's can be decelerated into the low energy region by this simplified system, obtaining beam for rare decay at ~0.04 μ/p .

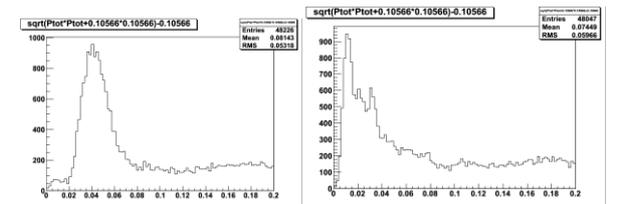

Figure 7: Beam energy distributions at the end of the rotator and after deceleration.

The muons could also be decelerated by passing through an absorber, with the thickness of the absorber tuned to maximize the number of μ's within the rare decay experiment acceptance window. This was attempted by adding an absorber immediately following the Rotator of fig. 4. An ~15cm LiH absorber appeared to be an initial optimum. Energy loss in an absorber appeared to be less efficient than systematic rf as a decelerator, since the absorber increased the beam energy spread, and the acceptance into a low-energy experiment would be reduced by a factor of 2 or more. (~0.02 μ/p)

However, an absorber is cheaper than the equivalent decelerating rf and could be used in an initial lower cost experimental configuration.

The efficiency of energy-loss absorption could be improved by introducing dispersion and adding a wedge component so that higher-energy μ's pass through more material. Absorbers tailored to the beam energy distribution could greatly increase the delivery into an energy acceptance window.

## ANOTHER FRONT END

The rotator is somewhat inefficient in moving the bunches to equal energies, and a more compact system can be developed. In an initial scenario the high-frequency buncher is followed by an 8m system, matched to place bunches at nearly equal momentum. (~120 MeV/c) This is followed by 4.2m with 6 cavities that decelerate the muons to ~50 MeV/c, which is matched to the needs of future mu2e experiments [7].

Simulations show that a very large number of μ's are moved to the desired energy( see fig. 8-9). More generally, the exercise indicates that variation and fitting of individual cavity properties may provide a more efficient φ-E rotation and deceleration than the initial general algorithm of our first scenarios. These variations should be explored in future research.

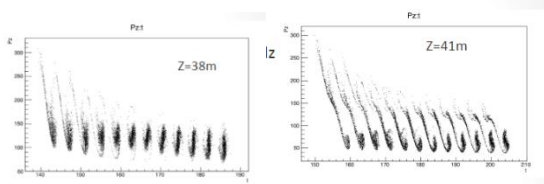

Fig. 8: $P_z$-$c\tau$ projections of beam after a simplified rotation using 10 rf cavities followed by deceleration to rare-decay experiment energies.

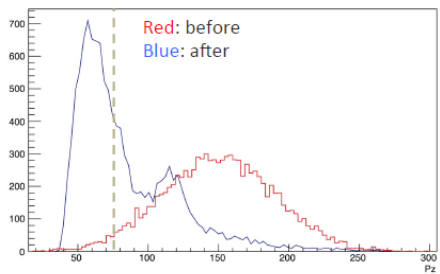

Figure 9: Momentum histogram of beam before and after the deceleration.

## VARIATIONS TO CONSIDER

The simulations presented here began with a MAP-based target and capture within a 15 T solenoid from an 8 GeV proton beam. A lower-energy proton beam would be more efficient at low energy μ capture, and a lower field capture lens should also be adequate in capturing low-momentum π; these variations will be explored, and optimized.

While not included in initial simulation, a complete design would include a bent-solenoid chicane in the initial drift, similar to that shown in figure 2. This would separate high-energy π's from the downstream μ buncher, obtaining a cleaner low-E μ beam. MAP studies show that the chicane would not reduce the low-E flux. Precise parameters will require optimization studies.

## CONCLUSION

These early explorations and simulations definitely show that the MAP front end design concepts can be adapted to obtain high quality low energy μ beams for future experiments. The simplified cases described here would be dramatically more affordable than the original HEP designs and should be within the budget of next-generation rare-decay experiments.